\documentclass[%
 reprint,
 amsmath,amssymb,
 aps,
]{revtex4-1}

\usepackage{siunitx}
\usepackage{graphicx}
\usepackage{dcolumn}
\usepackage{bm}

\begin{document}

\preprint{APS/123-QED}

\title{Experimental conversion of position correlation into polarization entanglement }

\author{Chithrabhanu Perumangatt}
\affiliation{Centre for Quantum Technologies, National University of Singapore, 3 Science Drive 2, S117543, Singapore}
\author{Alexander Lohrmann}%
\affiliation{%
Centre for Quantum Technologies, National University of Singapore, 3 Science Drive 2, S117543, Singapore
}%
\author{Alexander Ling}%
 
\affiliation{%
 Centre for Quantum Technologies, National University of Singapore, 3 Science Drive 2, S117543, Singapore
}%
\affiliation{%
 Department of Physics, National University of Singapore, 2 Science Drive 3, S117542, Singapore
}%

\begin{abstract}


This manuscript presents a method to convert position correlation of photon-pairs into  polarization entanglement. This is achieved by individually manipulating the polarization state of photons generated in different parts of a non-linear medium and putting them in coherent superposition. This concept is experimentally demonstrated using photon-pairs produced by spontaneous parametric down conversion (SPDC). The method was used to implement a compact source producing an observed photon-pair rate of 120,000/s/mW with an entanglement fidelity of 0.99. This method can be extended to any photon-pair generation process with initial position correlation. 

\end{abstract}

\pacs{Valid PACS appear here}
\maketitle


Photons entangled in polarization have been a prime tool for experiments in fundamental quantum physics~\cite{teleport_experimenetal, DenseCoding_experimental, EPR_experimental, Swaping_exp,DelayedChoice_exp}, and they are an essential part of many quantum communication schemes~\cite{QC144KM2007,143Teleport,micius_EQKD}. One of the most mature methods to create such photons is using spontaneous parametric down conversion (SPDC), where a pump photon splits into two lower energy photons under conservation of energy and momentum in a nonlinear crystal. 
 To  generate polarization entanglement from SPDC, it is necessary to impose coherent superposition of two  pump-decay paths. 
 
 There is a variety of techniques by which this coherent superposition can be achieved.  Examples of previous entangled photon-pair sources have utilized  momentum correlation ~\cite{Kwiat_typeII}, indistinguishable  pump decay in two separate crystals~\cite{ Kwiat98,Trojek,Steinlechner:12,Villar:18}, or two distinguishable pump beams~\cite{ SagnacTypeII,narrow_typeII,Steinlechner:14, fiorentino2008compact,NIST-loopholefree,broadband}.   
 The correlation in the birth position of the photon-pairs \cite{Schneeloch_2016} is yet another property that can potentially be exploited for the generation of polarization entanglement.  
Position correlation is  an intrinsic property of photon-pairs produced in any parametric process in a bulk medium. This position correlation is independent of the phase matching used in SPDC and is preserved unless it is filtered into a single spatial mode. Position and momentum correlations can be exploited
for higher dimensional entanglement and are used for sensing and imaging applications~\cite{Padgett_position, PhysRevLett.92.233601,PhysRevLett.94.220501}.

This manuscript reports the use of position correlation to implement high quality polarization entangled photon-pair sources.    
In this method, the polarization state of the photons produced at corresponding transverse positions in the non-linear crystal are manipulated before undergoing coherent superposition. A proof-of-concept demonstration is first performed by implementing a modified Mach-Zehnder interferometer for the manipulation and superposition. Next, the technique was applied to implement a bright entangled photon-pair source with minimum elements in a simple geometry. This technique can also be extended to other pair production processes such as four wave mixing in an atomic vapour \cite{HPS-4wavemixing,Atomic_vapor_HPS}.  

\begin{figure*}
 \centering
    \includegraphics{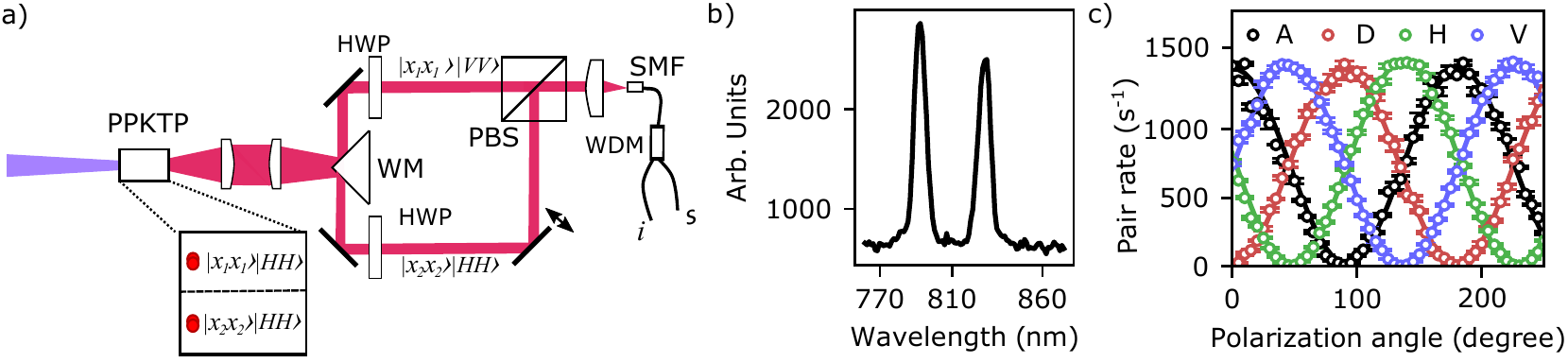}
    \caption{ a) Experimental setup  for the conversion of position correlation to polarization entanglement. The zoomed-in image of the SPDC interaction volume describes photon-pairs generated at the upper part of the crystal to be having  position state $\vert x_1, x_1\rangle$, while photon-pairs generated on the lower part are marked as $\vert x_2, x_2\rangle$. Two plano-convex lenses (f=15mm) are used to image the interaction volume onto a wedge mirror (WM). The photons are coupled into a single mode fiber (SMF) and a wavelength division multiplexer (WDM) is used to split the signal (s) and idler(i) photons. A mirror inside the interferometer is attached to a piezo-actuator to lock the phase while observing the constructive or distructive interference fringe produced by the pump laser.  b) Measured spectrum of the SPDC photons. 
    c) The two photon polarization correlation curves of the photon-pairs demonstrating high visibility. }
    \label{fig:1}
\end{figure*}

 Consider collinear type-0 SPDC ($H\rightarrow HH$) in a periodically poled crystal where the pump, signal and the idler photons are co-polarized.
The birth position of a photon-pair can be anywhere within the interaction volume. These positions are discretized into two bins by labels $x_1$ and $x_2$,  (see the zoomed-in image of the SPDC interaction volume in Fig.~\ref{fig:1}(a)) according to their origin  with respect to a conceptual line drawn through the centre of the interaction volume. The states generated in each discrete space can be written as
\begin{equation}
        \vert\phi_1\rangle =  \vert x_1, x_1\rangle\otimes\vert HH\rangle, \  
    \vert\phi_2\rangle =  \vert x_2, x_2\rangle\otimes\vert HH\rangle
\end{equation}
where $1$ and $2$ correspond to the different transverse positions, and $\vert HH \rangle$ indicates the horizontal polarization state of the photons. A polarization transformation $\vert HH\rangle \rightarrow \vert V V\rangle$ is applied to photons with the state $\vert\phi_1\rangle$ ($\vert\phi_1\rangle \rightarrow \vert\phi_1^\prime\rangle$ ).  The states $\vert\phi_1^\prime\rangle$ and $\vert\phi_2\rangle$ are coherently superposed via a polarizing beam splitter (PBS). The resultant state can be written as
\begin{equation}
    \vert\phi\rangle = \frac{1}{\sqrt{2}}\left( \vert x_2, x_2\rangle\otimes\vert H H\rangle+e^{i\Delta\varphi} \vert x_1 x_1\rangle\otimes\vert V V\rangle\right)
\end{equation}
 The photons are then projected into a single mode  erasing any position distinguishability. The state after the single mode projection is given by
\begin{equation}
        \vert\phi\rangle = \frac{1}{\sqrt{2}}\left( \vert H H\rangle+e^{i\Delta\varphi} \vert V V\rangle\right).
\end{equation}
By adjusting the constant phase offset, $\Delta\varphi$, the maximally entangled Bell states, $\vert\Phi^{\pm}\rangle$, can be generated.

 An implementation of the concept using a Mach-Zehnder interferometer is shown in Fig.~\ref{fig:1}(a). A laser beam of 405 nm with 160 MHz line-width is focused (pump waist of 150~$\si{\micro\metre}$) into a periodically poled potassium titanyl phosphate (PPKTP) crystal with a length of 10~mm, kept in an oven to regulate the temperature. Non-degenerate, collinear type-0 phase matching is employed for the generation of photon-pairs. The non-degenerate condition is used since the final separation of the signal and idler photons are done by wavelength.  The crystal temperature is tuned to generate signal and idler wavelengths at 792~nm and 829~nm, respectively as shown in Fig.~\ref{fig:1}(b). For exploiting the position correlation, the centre of the crystal (SPDC interaction volume)  is imaged onto a wedge mirror using two plano-convex lenses in a 4f configuration \cite{PhysRevLett.94.220501,Padgett_position}. The effective collection waist at the centre of the crystal is 75~$\si{\micro\metre}$. The wedge mirror splits the SPDC photons according to the birth position. 
 
 The polarization state of the photons reflected to the upward path  is converted from horizontal to vertical using a half wave plate (HWP). The polarization of the photons reflected to the downward path are not rotated. Nevertheless, they pass through a HWP oriented at 0$^\circ$ for matching the dispersion experienced by photons in the upper path.  Both beams are combined using a polarizing beam splitter and then coupled to a single mode fiber (SMF). The signal and idler photons are then separated using a wavelength division multiplexer (WDM).   The phase difference between horizontally and vertically polarized photon-pairs, before reaching the PBS is $\Delta\varphi=  2\pi\Delta L(\frac{1}{\lambda_s}+\frac{1}{\lambda_i})=2\pi\Delta L(\frac{1}{\lambda_p})$, where $\lambda_p $ is the pump wavelength. To get a 0 or $\pi$ phase, the interferometer is locked on to either the constructive or destructive interference fringe produced by the pump beam.

The photons exiting the WDM are detected on Geiger mode avalanche photo diodes (SAP500) with efficiencies of approximately 50\%. The observed pair rate normalized to pump power is 260~k/s/mW with pair-to-singles of 20\% and 16\% for signal and idler photons respectively. The polarization state of the signal and idler photons are analyzed for their non-local polarization correlations. The photon-pair polarization correlation curves are given in Fig.~\ref{fig:1}(c), and exhibit an average visibility of 0.995 with a corresponding \cite{fidality_from_visibility} fidelity of 0.997 towards the maximally entangled $\vert\Phi^+\rangle$ state.

The quality of the state depends upon the imaging system. In a non-ideal imaging system, for example when the wedge mirror is displaced from the focal position of the second lens, some of the SPDC photons will be separated by momentum, rather than position. 
This can introduce polarization anti-correlation resulting in a reduction in entanglement fidelity. Shifts in the transverse position of the wedge mirror can also result in a non-maximally entangled state. 
 
This interferometer setup could be used to generate anti-correlated $\vert\Psi^{\pm}\rangle$ Bell states. This can be done by utilizing momentum correlation and is achieved when a 2f lens system is used. This method splits the photon pair; instead of two photons going into either arm, the twin photons are separated at the wedge mirror and traverse each arm singly.
With this setup, however, it is difficult to achieve high fidelity because the phase dependence is now proportional to $\Delta L (\frac{1}{\lambda_s} - \frac{1}{\lambda_i})$.
Since the signal and idler wavelengths intrinsically have a broad bandwidth as shown in Fig.~\ref{fig:1}(b), (unless strong wavelength filtering is used, which in turn has a strong impact on brightness), generation of high quality $\vert\Psi^{\pm}\rangle$ states require the absolute path length difference in the current implementation to be less than \SI{20}{\micro\metre}.
 
\begin{figure}[t]
 \centering
    \includegraphics{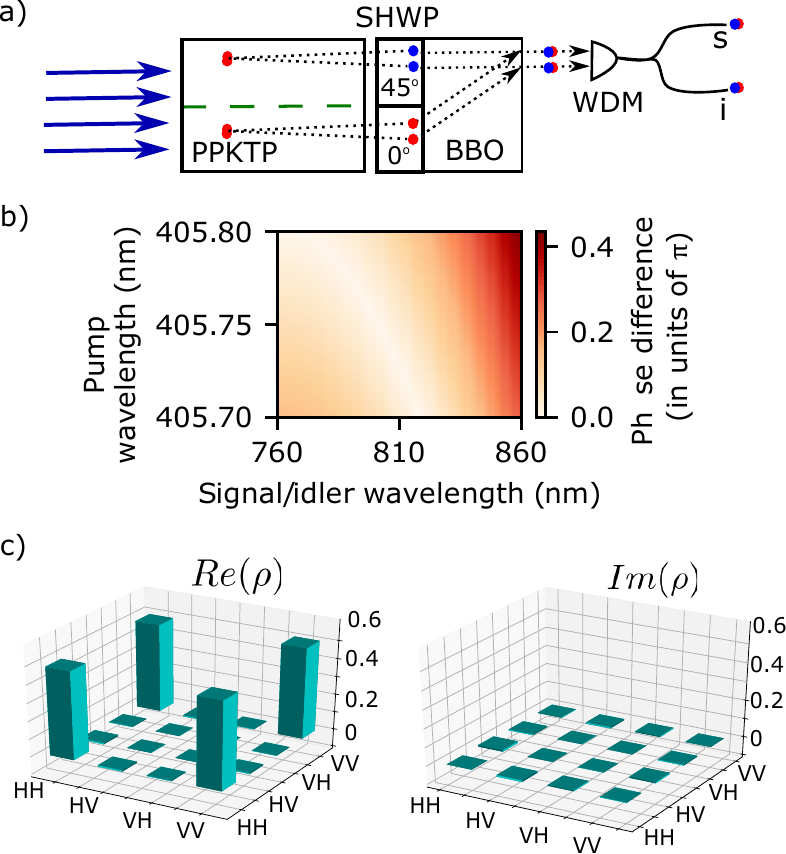}
    \caption{ a) Schematic of the compact polarization entangled photon source. The dashed line represent the distinction between photons generated at different parts of the PPKTP crystal. Red and blue filled circles represent horizontally and vertically polarized photons respectively. The WDM separates the photon-pairs according to its wavelength   b) The phase difference between $\vert HH\rangle $ and $\vert VV\rangle $ states for different pump and signal wavelengths. c) Real and imaginary parts of the density matrix for the state generated using the compact setup.  }
    \label{fig:2}
\end{figure}

The inherent instability of the interferometer makes it challenging to deploy the source in a rugged environment \cite{Tang2016}. A design that is intrinsically more stable is shown in Fig.~\ref{fig:2}(a). 
In this approach the PPKTP crystal is pumped with a collimated beam and the generated SPDC photon-pairs undergo polarization treatment as soon as they exit the crystal. 
The photon-pairs, after undergoing polarization manipulation, are coherently superposed using the walk-off in a nonlinear crystal and coupled into a single mode fiber.

In the actual implementation, a collimated pump beam with a waist size of approximately \SI{400}{\micro\metre} is used. The SPDC photons pass through an achromatic segmented half wave plate (SHWP) placed at the exit face of the PPKTP crystal. One half of the SHWP is oriented at 45 degrees while the other half is oriented at 0 degrees. The use of a segmented half-wave plate is convenient since the total phase acquired in both paths will be approximately equal. 

 To combine the two paths, the source uses a Beta Barium Borate (BBO) crystal with a cut angle of $28.8^\circ $ and a length of 4~mm. The length of the nonlinear crystal is chosen such that it introduces a displacement that is half the pump beam width. The superposed photons are then collected into a single mode fibre before being separated according to their wavelength. The effective collection waist at the centre of the crystal is 75~$\si{\micro\metre}$. 
 
 The use of spatial walk-off for the coherent superposition of SPDC from two separate co-propagating pump beams have been used in previous work for the generation of polarization entanglement \cite{fiorentino2008compact,NIST-loopholefree,broadband}. 
 A major difference of this work from previous reports is that the source under discussion requires only a single pump beam and the two decay paths are generated conceptually from a single SPDC interaction volume.   

Compared to the free space interferometer, there is a wavelength dependent phase (Fig.~\ref{fig:2}(b)) introduced by the BBO crystal which has a chromatic birefingence. In principle, it could be corrected using  a positive uniaxial crystal, e.g. Yttrium Orthovanadate (YVO$_4$). However,  the SPDC spectrum is sufficiently narrow (see Fig.~\ref{fig:1}(b)) and the corresponding phase  is negligible.  The constant phase $\Delta\varphi$, due to the relative path length difference between $\vert HH\rangle $ and $\vert VV\rangle $ states, is corrected by tilting the BBO crystal. One could also tune this constant phase actively by introducing a liquid crystal retarder~\cite{Lohrmann:19}.

With the compact source, the observed brightness is 120~k/s/mW with pair-to-singles ratios of 18\% and 15\% for signal and idler, respectively. Quantum state tomography on the generated state shows that the fidelity towards the $\vert\Phi^+\rangle$ state is 0.991 (see Fig.~\ref{fig:2}(c)).

Brightness and pair-to-singles ratios are reduced compared to the larger interferometer-based design which is attributed to scattering losses at the interface of the two parts of the SHWP. This is difficult to overcome since the interface overlaps with the maximum intensity of the pump beam. This also prevents the use of a strongly focused pump. Improved tolerances in the fabrication of the SHWP should enable a more focused pump, which could improve the coupling of the SPDC emission into the selected collection mode, thus improving the brightness of the source.

  This source is a good candidate for constructing rugged and  compact entangled photon-pair source. The same method can be applied for generating entanglement in telecommunication wavelengths as well. Position correlation is independent of spectral and spatial coherence~\cite{coherence_positioncorr}, permitting the use of a broadband pump~\cite{broadband} which will be investigated in future efforts. The necessary and sufficient requirement for our method to generate polarization entanglement are photon-pairs with initial position correlation. This condition can be found in various processes, such as SPDC in bulk crystals or four-wave mixing in atomic vapours~\cite{HPS-4wavemixing,Atomic_vapor_HPS, Gulati_2015}. Applying this effect in atomic vapours could provide another route to observing polarization entangled photons with line-widths compatible with atomic systems. In conclusion, we have demonstrated that the inherent position correlation of the photon-pairs generated via SPDC in bulk media is another route for the implementation of compact, high performing polarization entangled photon-pair sources. 

This research is supported by  National  Research  Foundation  (NRF), Award No.  R-710-00-133-281 and the Singapore Ministry  of  Education  (MOE)  Research  Centres of Excellence (RCE) program.

\end{document}